\documentclass[sigconf, anonymous=False]{acmart}
\usepackage{algorithmic}
\usepackage{graphicx}
\usepackage{textcomp}
\usepackage{xurl}
\usepackage{float}
\usepackage{diagbox}
\usepackage{subcaption}
\usepackage{booktabs}
\usepackage{multicol}
\usepackage{hyperref}
\usepackage{cuted}
\usepackage{makecell}
\usepackage{xspace}
\usepackage{listings}

\copyrightyear{2023} 
\acmYear{2023} 
\setcopyright{acmlicensed}\acmConference[CF '23]{20th ACM International Conference on Computing Frontiers}{May 9--11, 2023}{Bologna, Italy}
\acmBooktitle{20th ACM International Conference on Computing Frontiers (CF '23), May 9--11, 2023, Bologna, Italy}
\acmPrice{15.00}
\acmDOI{10.1145/3587135.3592211}
\acmISBN{979-8-4007-0140-5/23/05}

\colorlet{punct}{red!60!black}
\definecolor{background}{HTML}{EEEEEE}
\definecolor{delim}{RGB}{20,105,176}
\colorlet{numb}{magenta!60!black}

\lstdefinelanguage{json}{
    basicstyle=\footnotesize\ttfamily,
    numbers=left,
    numberstyle=\scriptsize,
    stepnumber=1,
    numbersep=5pt,
    showstringspaces=false,
    breaklines=true,
    frame=single,
    xleftmargin=2em,
    xrightmargin=2em,
    literate=
     *{:}{{{\color{punct}{:}}}}{1}
      {,}{{{\color{punct}{,}}}}{1}
      {\{}{{{\color{delim}{\{}}}}{1}
      {\}}{{{\color{delim}{\}}}}}{1}
      {[}{{{\color{delim}{[}}}}{1}
      {]}{{{\color{delim}{]}}}}{1},
}

\lstdefinelanguage{python}{
    basicstyle=\tiny\ttfamily,
    numbers=left,
    numberstyle=\scriptsize,
    stepnumber=1,
    numbersep=5pt,
    showstringspaces=false,
    breaklines=true,
    frame=single,
    xleftmargin=2em,
    xrightmargin=2em,
    literate=
     *{:}{{{\color{punct}{:}}}}{1}
      {,}{{{\color{punct}{,}}}}{1}
      {\{}{{{\color{delim}{\{}}}}{1}
      {\}}{{{\color{delim}{\}}}}}{1}
      {[}{{{\color{delim}{[}}}}{1}
      {]}{{{\color{delim}{]}}}}{1},
}

\lstdefinelanguage{yaml}{
    basicstyle=\Tiny\ttfamily,
    numbers=left,
    numberstyle=\scriptsize,
    stepnumber=1,
    numbersep=5pt,
    showstringspaces=false,
    breaklines=true,
    frame=single,
    xleftmargin=2em,
    xrightmargin=2em,
     string=[s]{'}{'},
     stringstyle=\color{blue},
     comment=[l]{:},
     commentstyle=\color{black},
     morecomment=[l]{-}
 }

\newcommand{\ff}{\textit{FastFlow}\xspace}

\definecolor{azure}{rgb}{0, 0.28, 0.67}
\newcommand{\new}[1]{{{~#1}}}

\newif\ifanonymize
\anonymizefalse

\ifanonymize
\newcommand{\db}[1]{REDACTED}
\else
\newcommand{\db}[1]{#1}
\fi

\def\BibTeX{{\rm B\kern-.05em{\sc i\kern-.025em b}\kern-.08em
    T\kern-.1667em\lower.7ex\hbox{E}\kern-.125emX}}
\begin{document}

\title{\new{Experimenting with Emerging RISC-V Systems for Decentralised Machine Learning}}


\author{Gianluca Mittone}
\email{gianluca.mittone@unito.it}
\orcid{0000-0002-1887-6911}
\affiliation{
  \institution{University of Turin}
  \city{Turin}
  \country{Italy}
}
\author{Nicoló Tonci}
\email{nicolo.tonci@phd.unipi.it}
\orcid{}
\affiliation{
  \institution{University of Pisa}
  \city{Pisa}
  \country{Italy}
}
\author{Robert Birke}
\email{robert.birke@unito.it}
\orcid{}
\affiliation{
  \institution{University of Turin}
  \city{Turin}
  \country{Italy}
}
\author{Iacopo Colonnelli}
\email{iacopo.colonnelli@unito.it}
\orcid{}
\affiliation{
  \institution{University of Turin}
  \city{Turin}
  \country{Italy}
}
\author{Doriana Medić}
\email{doriana.medic@unito.it}
\orcid{}
\affiliation{
  \institution{University of Turin}
  \city{Turin}
  \country{Italy}
}
\author{Andrea Bartolini}
\email{a.bartolini@unibo.it}
\orcid{}
\affiliation{
  \institution{University of Bologna}
  \city{Bologna}
  \country{Italy}
}
\author{Roberto Esposito}
\email{roberto.esposito@unito.it}
\orcid{}
\affiliation{
  \institution{University of Turin}
  \city{Turin}
  \country{Italy}
}
\author{Emanuele Parisi}
\email{emanuele.parisi@unibo.it}
\orcid{}
\affiliation{
  \institution{University of Bologna}
  \city{Bologna}
  \country{Italy}
}
\author{Francesco Beneventi}
\email{francesco.beneventi@unibo.it}
\orcid{}
\affiliation{
  \institution{University of Bologna}
  \city{Bologna}
  \country{Italy}
}
\author{Mirko Polato}
\email{mirko.polato@unito.it}
\orcid{}
\affiliation{
  \institution{University of Turin}
  \city{Turin}
  \country{Italy}
}
\author{Massimo Torquati}
\email{massimo.torquati@unipi.it}
\orcid{}
\affiliation{
  \institution{University of Pisa}
  \city{Pisa}
  \country{Italy}
}
\author{Luca Benini}
\email{luca.benini@unibo.it}
\orcid{}
\affiliation{
  \institution{University of Bologna}
  \city{Bologna}
  \country{Italy}
}
\author{Marco Aldinucci}
\email{marco.aldinucci@unito.it}
\orcid{}
\affiliation{
  \institution{University of Turin}
  \city{Turin}
  \country{Italy}
}

\renewcommand{\shortauthors}{Mittone et al.}

\begin{abstract}
Decentralised Machine Learning (DML) enables collaborative machine learning without centralised input data. 
Federated Learning (FL) and Edge Inference are examples of DML. 
While tools for DML (especially FL) are starting to flourish, many are not flexible and portable enough to experiment with novel processors (e.g., RISC-V), non-fully connected network topologies, and asynchronous collaboration schemes. 
We overcome these limitations via a domain-specific language allowing us to map DML schemes to an underlying middleware, i.e. the \ff parallel programming library.
\new{We experiment with it by generating different working DML schemes on x86-64 and ARM platforms and an emerging RISC-V one.}
We characterise the performance and energy efficiency of the presented schemes and systems. As a byproduct, we introduce a RISC-V porting of the PyTorch framework, the first publicly available to our knowledge. 
\end{abstract}



\pagenumbering{gobble}

\begin{center}
The following paper is the accepted version of ACM copyrighted material
\\[12pt]
\textit{Gianluca Mittone, Nicoló Tonci, Robert Birke, Iacopo Colonnelli, Doriana Medić, Andrea Bartolini, Roberto Esposito, Emanuele Parisi, Francesco Beneventi, Mirko Polato, Massimo Torquati, Luca Benini, and Marco Aldinucci. 2023. Experimenting with Emerging RISC-V Systems for Decentralised Machine Learning. In Proceedings of the 20th ACM International Conference on Computing Frontiers (CF '23). Association for Computing Machinery, New York, NY, USA, 73–83}
\\[12pt]
presented at the CF'23 conference in Bologna, Italy.
\\[12pt]
DOI: \href{https://doi.org/10.1145/3587135.3592211}{10.1145/3587135.3592211}
\end{center}

\maketitle

\pagestyle{plain}
\section{Introduction}
\label{sec:intro}

Recent years have been characterised by crucial advances in Machine Learning (ML) systems.
These advancements have been made possible thanks to the widespread availability of massive and ubiquitous computational resources and copious and distributed data sources.
The consequent deployment of ML methods across many industries has generated concerns about data access and movement, such as performance, energy efficiency, security, and privacy~\cite{goodman2017european,pardau2018california,chen2021understanding}.
Furthermore, companies consider collected data as competing advantages and are unwilling to share it outside the organisation. This results in data being dispersed into isolated islands, and ML practitioners are forbidden from collecting, fusing, and ultimately using the data to improve their systems.


Decentralised ML (DML) using Federated Learning (FL) and Edge Inference (EI) tackles the difficulties mentioned above, but practical implementations are not straightforward.
\new{The emergence of alternative ISAs to x86-64 and ARM-v8, such as RISC-V, exacerbates the challenges in porting (optimised) libraries and guaranteeing interoperability between heterogeneous systems.}
Many off-the-shelf frameworks are available, each specifically designed to implement one of a few use-case scenarios.
First, while standard FL is based on a master-worker approach, few emerging techniques explore distributed, sparse graphs.
Second, to our knowledge, no FL framework still allows the user to specify a personalised, experimental communication graph between the parties taking part in the federation, posing a severe limit to researchers.
Third, most current FL frameworks are implemented in Python, and their communication infrastructure is based on the gRPC and protobuf technologies.
While this approach is practical, we advocate that it is not the most efficient.
Finally, no FL framework has yet been proven on the RISC-V ISA. 
A modern DML framework should be flexible in defining the system architecture and communication patterns to open up research on alternative FL systems and be able to exploit optimised distributed runtimes for optimal performance on different ISAs.


Given this objective, we propose a top-down methodology to describe and implement flexible and fast DML systems. 
Our methodology leverages two proven tools: a formal language designed for describing parallel processes, namely RISC-pb$^2$l (RISC Parallel Building Block Library), which we adapt to the DML domain, and a lightweight C++ distributed run-time, namely \ff{}, which we port to the emerging RISC-V hardware.
The translation between these two components is effortless since any valid DML system description can be directly translated into a \ff program. 
\new{We demonstrate the effectiveness and flexibility of our approach by showcasing it across many experiments on three different learning schemes, spanning FL and EI, and different hardware architectures, x86-64, ARM and the emerging RISC-V, characterising their performance and energy efficiency.}

In summary, our contributions are as follows:
\begin{itemize}
    \item a methodology to design flexible, decentralised learning schemes and implement them on a
    \emph{lightweight distributed runtime} enabling the experimentation and operation of ML at heterogeneous edge, especially aiming at FL and EI;
    \item proving system design flexibility via three different FL and EI use cases (master-worker, peer-to-peer and tree-based aggregation);
    \item\new{an evaluation to showcase support of x86-64, ARM and emerging RISC-V ISAs;}
    \item the first publicly available porting of the PyTorch framework and \ff library for RISC-V.
\end{itemize}

\begin{table*}[ht]
\caption{Brief overview of some of the mature FL frameworks available on the market, based on~\cite{beltran2022decentralized}.\label{tab:frameworks}}
{\small
\begin{tabular}{lllllc}
\toprule
{\bf Framework} & {\bf Target} & {\bf Scenario} & {\bf Comm. protocol} & {\bf Implementation} & \makecell[l]{\bf Programmable\\ \bf comm. graph} \\ 
\midrule
TFF~\cite{tff}             & Cross-silo                            & Simulation/Real & gRPC/proto                                    & Python            & \textcolor{red}{\huge\texttimes}      \\ \midrule
PySyft~\cite{ziller2021pysyft}         & Cross-silo/Cross-device & Simulation/Real & Websockets                                    & Python            & \textcolor{red}{\huge\texttimes}      \\ \midrule
SecureBoost~\cite{cheng2021secureboost}        & Cross-silo                            & Simulation/Real & gRPC/proto                                    & Python            & \textcolor{red}{\huge\texttimes}      \\ \midrule
FederatedScope~\cite{xie2022federatedscope}     & Cross-silo/Cross-device & Simulation/Real & gRPC/proto                                    & Python            & \textcolor{red}{\huge\texttimes}      \\ \midrule
LEAF~\cite{caldas2018leaf}                & Cross-silo                            & Simulation                    & -                                             & Python            & \textcolor{red}{\huge\texttimes}      \\ \midrule
FedML~\cite{he2020fedml}          & Cross-silo                            & Simulation/Real & \makecell[l]{gRPC/proto \\ MPI, MQTT}          & Python            & \textcolor{red}{\huge\texttimes}      \\ \midrule
OpenFL~\cite{reina2021openfl}              & Cross-silo                            & Simulation/Real & gRPC/proto                                    & Python            & \textcolor{red}{\huge\texttimes}      \\ \midrule
Flower~\cite{beutel2020flower}              & Cross-silo/Cross-device & Simulation/Real & gRPC/proto                                    & Python            & \textcolor{red}{\huge\texttimes}      \\ \midrule
Our proposal        & Cross-silo                           & Simulation/Real & \makecell[l]{\bf TCP/Cereal \\ \bf MPI/Cereal}   & \textbf{C/C++}    & \textcolor{green}{\large\checkmark}   \\ \toprule
\end{tabular}
}
\end{table*}

\section{Background and Motivation}
\label{sec:dml}

\subsection{Federated Learning}
Federated learning has been proposed by~\cite{mcmahan2017communication} as a way to develop better ML systems without compromising the privacy of final users and the legitimate interests of private companies. 
Initially deployed by Google for predicting text input on mobile devices, FL has been adopted by many other industries, such as mechanical engineering and health care~\cite {li2020review}.

FL is a learning paradigm where multiple parties (\emph{clients}) collaborate in solving a learning task using their private data. 
Importantly, each client's data is not exchanged or transferred to any participant. 
Clients collaborate by exchanging local models via a central server in its most common configuration. 
The server (\emph{aggregator}) collects and aggregates the local models to produce a global model. 
The global model is then sent back to the clients, who use it to update their local models. 
Then, using their private data, they further update the local model. 
This process is repeated until the global model converges to a satisfactory solution or another termination condition is met (e.g., a maximum number of rounds). 

There are two main FL settings: cross-device and cross-silo, with different challenges.
In cross-device FL, the parties can be edge devices (e.g., smart devices and laptops), and they can be numerous (order of thousands or even millions). 
Parties are considered not reliable and with limited computational power.
In the cross-silo FL setting, the involved parties are organisations; the number of parties is limited, usually in the $[2, 100]$ range. 
Given the nature of the parties, communication and computation are no real bottlenecks.

\subsection{Edge Inference}
Inference in DNNs~\cite{wu2019machine} is much more lightweight than training. \new{Notice that, roughly speaking, one may expect the single inference step cost to be about $1/3$ of the cost needed for performing a single training step of the same model on similarly sized data due to additional loss evaluation and backward pass required in the training phase~\cite{adolf16fathom, li2020pytorch}.}
Relative performance varies significantly between network architectures (e.g., convolutional networks tend to pay a higher cost during training) and between system architectures (e.g., executing the operations on GPU may change the relative cost of training and inference). The inference is generally computed on a single (possibly SIMD/GPU/TPU accelerated) processing element. \new{At the distributed system level, a distributed inference process can be described with a directed acyclic graph, whereas FL requires a cyclic graph. This makes both cases interesting for experimentation.}

\subsection{Limits of Mainstream FL Frameworks}

Different FL frameworks already exist on the market~\cite{beltran2022decentralized}. However, they are relatively homogeneous, mainly targeting DNNs and focusing on the goodness of the learned model rather than the performance of the distributed infrastructure and communications involved in the federated process.

Table~\ref{tab:frameworks} compares the FL frameworks found mature by~\cite{beltran2022decentralized}.
Most of these target the cross-silo scenario, with only a few capable of handling the intricacies of large-scale, unreliable cross-device training, such as FederatedScope~\cite{xie2022federatedscope} and Flower~\cite{beutel2020flower}.
Similarly, all reported FL frameworks support a simulation mode, allowing experimenting and debugging a Federated system locally; however, it is not trivial that they also support a real-world-oriented distributed mode, like TFF~\cite{tff}, SecureBoost~\cite{cheng2021secureboost}, and OpenFL~\cite{reina2021openfl}.
From the abovementioned perspective, the most limited FL framework is LEAF~\cite{caldas2018leaf}: this software is explicitly designed to be used only for benchmarking purposes.
From a communication infrastructure perspective, almost all frameworks rely on the gRPC/proto combo.
While such an implementation is effective and well-supported, some implementations try to offer alternatives: this is the case of PySyft~\cite{ziller2021pysyft}, which offers a communication interface based on web sockets, and FedML~\cite{he2020fedml}, which supports a wide range of communication backends, including MPI and MQTT for, respectively, more high-performance and energy-efficient communication.

All listed FL frameworks share two characteristics, though: i) they are all implemented in Python, thus not taking performances seriously into account; and ii) all of them offer just one (or a small subset of) federation scheme; in other words, this software is highly specialised in handling just one communication graph and does not allow the user to specify and experiment with their personalised, complex, non-standard federation schema. 
These implementation choices constrain the possibility offered to researchers, who must roam over a vast range of FL frameworks to study how different federation schemes perform at the system and model levels.
Furthermore, since all FL frameworks have similar design ideas and implementation, there is no real alternative besides Python for FL practitioners.
This fact can be problematic, especially when working with experimental or embedded hardware architectures which are power and computation constrained, such as edge nodes using emerging RISC-V CPUs.
In contrast, our proposal leverages C/C++ for highly optimised code execution and a domain-specific language to specify the most suitable federation scheme.
\new{We verified experimentally the overhead introduced by using PyTorch's Python interface in place of the C++ LibTorch interface on the RISC-V platform.  
We chose RISC-V to stress the difference on a platform with low computational power and lack of optimised libraries.
%
Using the MNIST benchmark from the official PyTorch examples, we trained the same CNN, comparing the runtimes of the two APIs. 
The results averaged across 5 runs show that the C++ API (314.5s) is about 30\% faster than the Python API (442.8s).
}

\section{A top-down approach to FL}
\label{sec:tools}

We propose a top-down approach to implement a FL (or EI) system in the following.
\new{We show that modifying the common FL abstraction stack for more flexible and powerful frameworks is possible. This approach led us to deal with high-level issues, such as modelling of the distributed computation, and low-level issues, such as software porting to the RISC-V platform.}
The process starts with the definition of the system through a formal language for describing distributed processes, namely an adapted version of the RISC-pb$^2$l~\cite{aldinucci2014design} language.
This definition then translates into an implementation by mapping it to the building blocks of a computation middleware, namely the \ff parallel library~\cite{aldinucci2017fastflow}.
This workflow is straightforward and will be practically demonstrated in Section~\ref{sec:uc}.



\subsection{Describing the FL System}
The first step in creating our FL framework is to design a high-level modelling language. 
Such a language should be sufficiently abstract to allow to model a wide range of distributed computations while still being sufficiently specific to be easily applicable in different contexts.
With this aim in mind, we use the RISC-pb$^2$l formal language~\cite{aldinucci2014design} for parallel processes and adapt it to describe distributed FL workloads. 
RISC-pb$^2$l defines so-called Building Blocks (\texttt{BBs}), recurrent data-flow compositions of concurrent activities working in a streaming fashion, which are the primary abstraction layer for building parallel patterns and, more generally, streaming topologies~\cite{TorquatiPhD, aldinucci2014design}.
\new{Furthermore, RISC-pb$^2$l 
has an already available high-performance software implementation, that is \ff, which is also compatible with the RISC-V ISA. These facts were crucial for choosing RISC-pb$^2$l as language abstraction for this research work.}

We make RISC-pb$^2$l location-aware to suit the FL systems characteristics better since, in this scenario, each node owns different data. More specifically, we introduce a \emph{distribute} building block which computes distributively a function on a set of nodes specified via a superscript.
Table~\ref{tab:risc} reports the syntax and description of all RISC-pb$^2$l building blocks used in this paper. 
For the sake of space, we refer to~\cite{aldinucci2014design} for the complete set and detailed composition grammar.
The advantage of 
such a formalism is the possibility to design and discuss the distributed system before its implementation in a clear and structured way. 
Furthermore, 
this formalism allows discussion of the computational properties of a design 
and also to apply optimisations before the implementation phase.

\begin{table}[t]
\caption{Brief description of used RISC-pb$^2$l building block.\label{tab:risc}}
{\small
\begin{tabular}{ll}
\toprule
{\bf Syntax} & 
  {\bf Semantics} \\ \midrule
$\left(\left(f\right)\right)$ &
  \makecell[l]{\textbf{Seq wrapper} Wraps sequential code\\into a RISC-pb$^2$l “function”.} \\ \midrule
$\left(\left| f \right|\right)$ &
  \makecell[l]{\textbf{Par wrapper} Wraps any parallel code into\\a RISC-pb$^2$l “function”.} \\ \midrule
${\left[\left|\Delta\right|\right]}^N$ &
  \makecell[l]{\textbf{Distribute} Computes $|N|$ $\Delta$ distributively\\on the node set $N$ producing $|N|$ outputs.} \\ \midrule
$\Delta_1 \bullet \dots \bullet \Delta_n$ &
  \makecell[l]{\textbf{Pipe} Uses $n$ different programs as stages\\to process the input data items and to obtain\\output data items.} \\ \midrule
$\left(_g\vartriangleright \right)$ &
  \makecell[l]{\textbf{Reduce} Computes a output item using an $l$\\level ($l\ge1$) $k$-ary tree. Each node in the tree\\computes a $k$-ary function $g$.} \\ \midrule
$(_f\vartriangleleft)$ &
  \makecell[l]{\textbf{Spread} Computes $n$ output items using an\\$l$ level ($l\ge1$) $k$-ary tree. Each node in the\\tree computes a $k$-ary function $f$.} \\ \midrule
$\vartriangleleft_{D-Pol}$ &
  \makecell[l]{\textbf{1-to-N} Sends data received on the input\\channel to one or more output channels.\\$D-Pol\in\left[Unicast(p),Broadcast,Scatter\right]$.} \\ \hline
$\vartriangleright_{G-Pol}$ &
  \makecell[l]{\textbf{N-to-1} Sends data from the $n$ input\\channels on the single output channel.\\$G-Pol\in\left[Gather, Gatherall, Reduce\right]$.} \\ \midrule
$\overleftarrow{\left(\Delta\right)}_{cond}$ &
  \makecell[l]{\textbf{Feedback} Routes output data $y$ back to\\the input channel according to $Cond(x)$.}  \\ \bottomrule
\end{tabular}
}
\end{table}

\subsection{Implementing the FL System}
\label{sec:ff}

\begin{figure}[t]
    \centering

    \includegraphics[width=0.60\columnwidth,trim={0 0 0 0.2 cm},clip]{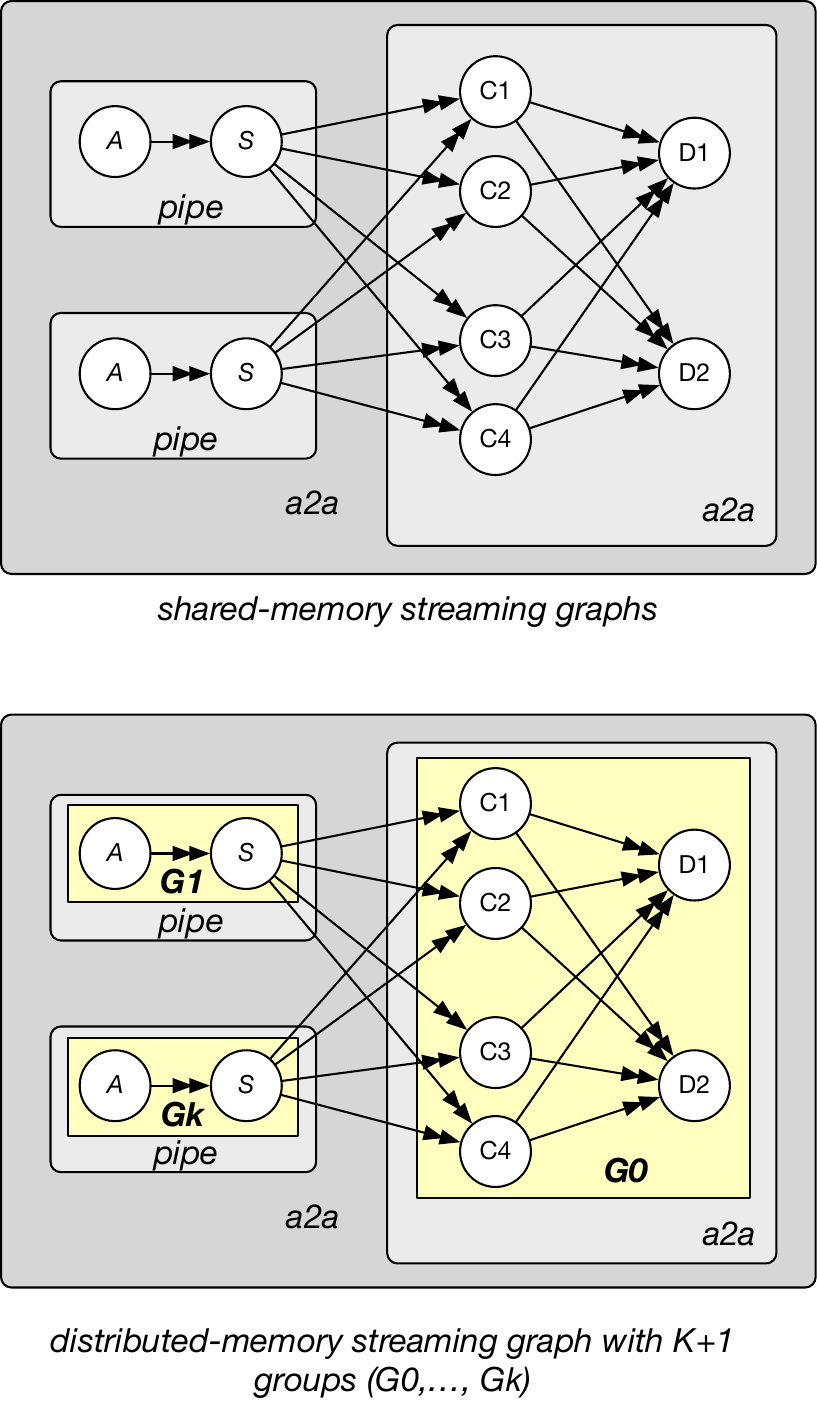}
    \caption{\textbf{Example of a \ff{} application:} The communication topology is described as a composition of building blocks in a data-flow graph, partitioned into distributed groups.}
    \label{fig:ffExampleGraph}
\end{figure}

We need to map the system description to runnable code to implement the FL (or EI) system. 
Here we use the C++ header-only \ff{} library~\cite{aldinucci2017fastflow}, which was developed alongside RISC-pb$^2$l. 
The \ff ~building blocks match in a one-to-one fashion the RISC-pb$^2$l building blocks, thus allowing for a straightforward implementation of a system given its formal specification.
Following the principles of the structured parallel programming methodology, a parallel application (or one of its components) is conceived by selecting and adequately assembling a small set of well-defined \texttt{BBs} modelling data and control flows.
These can be combined and nested in different ways, forming acyclic or cyclic concurrency graphs. 
The original \ff{} implements nodes as concurrent entities and edges as communication channels carrying data pointers. 
Recently, the \ff{} run-time system has been extended to deploy \ff{} programs in distributed-memory environments~\cite{Tonci2022DFF}. 
By introducing a small number of edits, the programmer may port shared-memory parallel \ff applications to a hybrid implementation (shared-memory plus distributed-memory) in which parts of the concurrency graph will be executed in parallel on different machines according to the well-known SPMD model.
Such refactoring involves introducing distributed groups (dgroups) which identify logical partitions of the \texttt{BBs} composing the application streaming graph according to a small set of graph-splitting rules. 
An example of a \ff application partitioned in distributed groups is given in Figure~\ref{fig:ffExampleGraph}.
Currently, inter-\textit{dgroup} (i.e., inter-process) communications leverage raw TCP/IP or MPI, whereas intra-\textit{dgroup} communications use highly efficient lock-free shared-memory communication channels~\cite{ffqueue-EuroPar2012}. 

The translation from RISC-pb$^2$l into \ff's BB is straightforward since BBs can be configured differently and composed together to obtain a one-to-one mapping.
In this sense, the \ff BB result to more expressive than their symbolical counterparts and can consequently be modelled to achieve efficient code implementations.
The expressiveness and flexibility of \ff's BB are offered in shared-memory and distributed-memory systems almost transparently. The translation has been done manually, but we envision that a small compiler would be capable of handling such a job.

\section{Use cases}
\label{sec:uc}

\begin{figure*}[t]
  \begin{center}
  \includegraphics[scale=0.45]{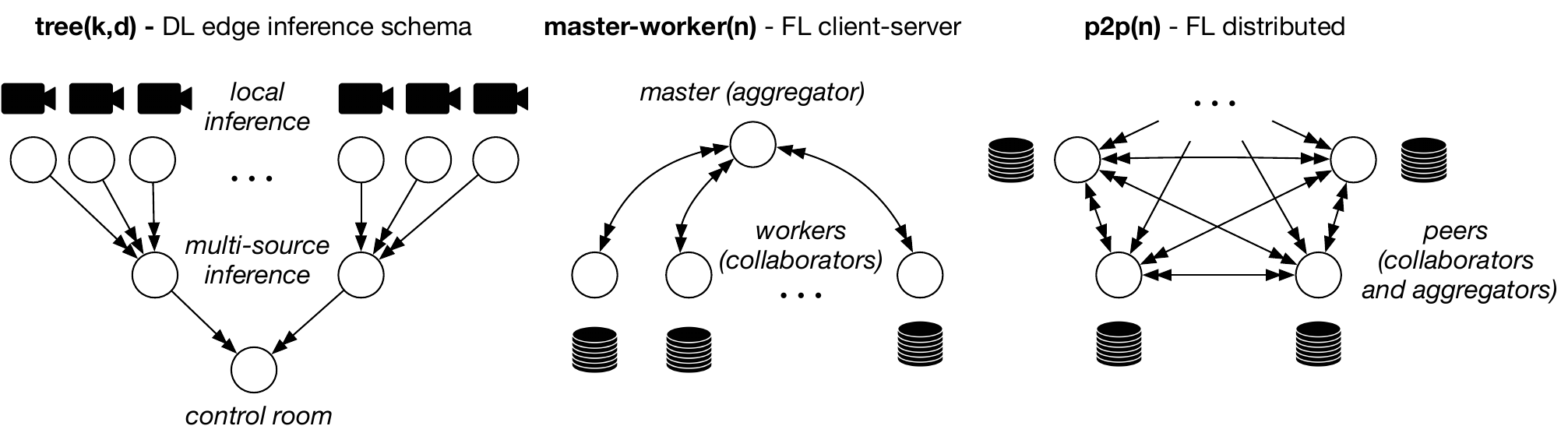}
  \caption{\textbf{The three considered use cases}: tree-based edge inference, master-worker, and peer-to-peer federated training.}
  \label{fig:schema}
  \end{center}
\end{figure*}


\subsection{Federated Learning}
We adopt federated averaging (FedAvg)~\cite{mcmahan2017communication} as aggregation strategy for our use cases since it is one of the most popular FL algorithms.
FedAvg organizes model training into rounds. 
At the start of each round, the aggregator distributes the weights of the aggregated (global) model to (a random subset of) participants; then, each client trains the model over one or more epochs using its local data before sending back the updated weights. 
Finally, the aggregator averages the received model weights to produce the updated global model for the next round.
To showcase the flexibility of our approach, we consider two system topologies for training (see Fig.~\ref{fig:schema}): a tree rooted at the aggregator mimicking the classic master-worker approach and a mesh made of peers, which combines the functionality of an aggregator and a worker, to avoid any central point of failure.

Using the RISC-pb$^2$l formalism, the master-worker FL process can be described as
\[
    {\small
    \left(\left(\text{init}\right)\right) \bullet 
    \overleftarrow{\biggl( 
    \left[\left|\left(\left|\text{test}\right|\right) \bullet 
    \left(\left|\text{train}\right|\right) \right|\right]^{W} \bullet
    \left(_{\text{FedAvg}}\vartriangleright\right) \bullet
    \vartriangleleft_{\text{Bcast}}
    \biggr)_{r}}}
\]
where $\left(\left(\text{init}\right)\right)$ is the function initialising the communication graph and creating the starting models, $W$ is the set of workers, and $r$ is the condition checking if the prefixed number of rounds has been reached.
On the other hand, the peer-to-peer scenario can be formalised through RISC-pb$^2$l as
\[
    {\small
    \left[\left|\left(\left(\text{init}\right)\right)\right|\right]^{P} \bullet
    \overleftarrow{\biggl( 
    \left[\left|\left(\left|\text{test}\right|\right) \bullet 
    \left(\left|\text{train}\right|\right) \bullet
    \vartriangleleft_{\text{Bcast}} \bullet
    \left(_{\text{FedAvg}}\vartriangleright\right) \right|\right]^{P}
    \biggr)_{r}}}
\]
As can be seen, the two formalisations are similar, which is not casual; it is possible to prove logically that the two formulas yield the same outputs if given the same inputs, modulo a different number of computations and communications.
This fact can be easily seen by comparing the last two terms of both feedback blocks; with a bit of rewriting, it is possible to state that
$(_{\text{FedAvg}}\vartriangleright) \bullet \vartriangleleft_{\text{Bcast}} \equiv [|\vartriangleleft_{\text{Ucast}_{A}}|]^{W} \bullet (_{\text{FedAvg}}\vartriangleright)$ and $[|\vartriangleleft_{\text{Bcast}} \bullet (_{\text{FedAvg}}\vartriangleright)|]^{P} \equiv [|\vartriangleleft_{\text{Bcast}}|]^{P} \bullet [|\vartriangleright_{\text{FedAvg}}|]^{P}$.
With this new formulation, it is easy to see that, even if they are equivalent output-wise, these two computations exploit different amounts of communications: ($[|\vartriangleleft_{\text{Ucast}_{A}}|]^{W}$ vs. $[|\vartriangleleft_{\text{Bcast}}|]^{P}$) and computations ($(_{\text{FedAvg}}\vartriangleright)$ vs. $[|\vartriangleright_{\text{FedAvg}}|]^{P}$).
\new{This result implies that the two computations are mathematically equivalent output-wise: this means that the two communication topologies will produce the same final ML model, with the same learning performances, assuming the same hyper-parametrisation and modulo the differences in communication involved in the process.}
This simple analysis exemplifies the potential of using a formal tool such as RISC-pb$^2$l for modelling and discussing distributed systems.


Given the two above descriptions, it is straightforward to translate them into \ff{} programs (serialisation is performed using the Cereal~\cite{grant2013cereal} library). 
Both topologies are based on the all-to-all \texttt{BB} (\texttt{ff\_a2a}), which efficiently models together the reduce and broadcast operators required by both workflows and rounds are modelled as cycles (i.e., feedback channels) created by activating the \texttt{wrap\_around} feature. 
The \texttt{ff\_a2a} \texttt{BB} defines two sets of nodes connected according to the shuffle communication pattern; this means that each node in the first set is connected to all nodes in the second set. 
If the first set contains the aggregator and the right set all worker nodes, then we have the master-worker topology.
On the other hand, for creating a mesh topology, peers are split into a modified aggregator, which additionally trains and includes a model trained on local data, and a distributor, which sends the local model to the other peers' aggregators; then, all aggregators are assigned to the left set and all distributors to the right set. 
Aggregator and distributor of the same peer are tied together by assigning them to the same distributed group. 
Leveraging the message routing options, we enforce that each aggregator sends its aggregated model only to its distributor; in contrast, each distributor forwards the aggregated model to all other aggregators on the feedback channel. 

Model training is based on the PyTorch library via the C++ API. 
All nodes are designed for modularity and implemented as derived \texttt{ff\_node} C++ classes. 
Multi-input multi-output nodes, i.e., aggregator node in tree topology and peer nodes in a mesh topology, are implemented exploiting the combiner \texttt{BB} (\texttt{ff\_comb}) by combining a multi-input adapter forwarding messages and a multi-output node containing the logic. Models are specified as the generic PyTorch \texttt{torch::nn::Module} class, allowing the framework to work with any PyTorch model. 
For future extensibility, the aggregation methodology is specified as a separate policy class, e.g., \texttt{FedAvg} for federated averaging. Similarly, workers allow specifying the training strategy, i.e., optimizer, to use as (automatic type inferred) template argument. 

The two FL use cases exploit a Multilayer Perceptron (MLP) made up of three fully connected layers trained to recognize digits from the MNIST dataset\footnote{http://yann.lecun.com/exdb/mnist/}; as hyper-parameters, we used cross-entropy as loss function and SGD as the optimizer, with a learning rate $0.01$ and momentum $0.5$. 
Concerning the MNIST dataset, we split the training set into equally sized random subsets assigned to each worker; this has been done to simulate a genuine federation in which each client possesses only a subset of the whole data distribution.


\subsection{Edge Inference}

The EI is showcased by using a tree topology for modelling a control-room use case, which aims to solve the underlying problem of raising alerts for man-on-the-ground events. 
Leaf nodes containing multiple cameras feed into local pre-trained standard YOLOv5 networks pre-processed video images by resizing them to the correct resolution (e.g., 640x640), adding a border, and ensuring that the image has the correct tensor shape. 
After applying the YOLOv5 model, the aggregation nodes post-process the result to extract the bounding boxes having a classification score larger than a given threshold (detecting man-on-the-ground events) and output aggregating results along the tree till the control room located at the root (see Fig.~\ref{fig:schema}). 
In our example, we modelled a three-level tree through RISC-pb$^2$l as

\[
    {\small
    \left(\left(\text{init}\right)\right) \bullet
    \overleftarrow{\biggl( 
    \left[\left|\text{infer}\right|\right]^{L} \bullet 
    \left(_\mathcal{F} \vartriangleright\right) \bullet 
    \left[\left|\text{combine}\right|\right]^{C} \bullet 
    \left(_\mathcal{F} \vartriangleright \right) \bullet 
    \left(\left(\text{alert}\right)\right)^{R} 
    \biggr)_{\infty}}}
\]

where $L, C, R$ are, respectively, the sets of leaf, control and root nodes, and $\mathcal{F}$ is the function routing to the father node.
The subsequent \ff{} implementation resulted in nesting two levels of all-to-all BB.

To work with PyTorch-based networks in a C++-only environment, we exported the network provided by the YOLOv5 maintainers into a TorchScript archive. 
TorchScript archives have the advantage of serialising the Python code describing the network and the necessary weights. 
The Python code is represented in the archive using a subset of Python itself (TorchScript is the name of this subset); weights are serialised in pickle format. 
When a TorchScript archive is deserialised, the Python is just-in-time compiled, and the binary is dynamically linked into the running program.


\section{Experimentation}

Each use case has been tested on three different hardware architectures hosted on the two research systems described below. 

\subsection{Computational infrastructures}

\subsubsection{Monte Cimone (RISC-V)}
The Monte Cimone~\cite{mcimone} system is the first physical prototype and test-bed of a complete RISC-V (RV64) compute cluster, integrating not only all the key hardware elements besides processors, namely main memory, non-volatile storage, and interconnect, but also a complete software environment for HPC, as well as a full-featured system monitoring infrastructure. 
Monte Cimone comprises eight computing nodes running Linux Ubuntu 21.04 and is enclosed in four computing blades. 
Each computing node is based on the U740 SoC from SiFive and integrates four U74 RV64GCB application cores, running up to 1.2 GHz and 16GB of DDR4, 1 TB node-local NVME storage, and PCIe expansion cards. 
Each Monte Cimone computing node integrates separated shunt resistors in series with each of the SiFive U740 power rails as well as for the on-board memory banks, which can be leveraged to attain fine-grained power monitoring of power rails, including the core complex, IOs, PLLs, DDR subsystem and PCIe one. 
The power rails current and voltage are monitored by a PXIe-4309 board from National Instruments. 
The PXIe-4309 module features 8 ADC devices and supports a maximum data acquisition rate of 2 MSamples per second. 
Collected current and voltage traces are then post-processed to obtain an average power consumption for each application run.

\subsubsection{EPI-TO (ARM-v8+x86-64+RISC-V)}
The EPI-TO system is a modular cluster designed to experiment with the technologies under development in the European Processor Initiative (EPI).
It includes an ARM-v8 module (4 nodes), an x86-64 module (4 nodes), and a RISC-V module (2 nodes). 
The x86-64 module comprises 4 Supermicro servers, each including 2 Intel Xeon Gold 6230 CPU (20-cores@2.10GHz) sockets, and 1536GB RAM. 
The ARM-v8 module comprises 4 ARM-dev kits, each including one socket Ampere-Altra Q80-30 (80-core@3GHz), 512GB RAM, 2 x NVidia BF-2 DPU, and 2 x NVidia A100 GPU. 
The Intel and Ampere servers are connected via an Infiniband HDR and a 1Gb/s Eth networks. 
The GPUs are not used in the present experiments. 
All nodes run Linux Ubuntu 20.04 and share a high-performance BeeGFS file system. 
All the servers are set to use the ``performance governor'' mode.
The RISC-V module is composed of 2 servers identical to Monte Cimone servers.

\begin{table*}[t]
\caption{Experimental setting description. FLOPs estimated using PyTorch profiler. In the case of YOLOv5n, only inference (forward pass) is performed. Data are reported when meaningful.\label{tab:setting}} 
    \centering
    {\small
    \begin{tabular}{llllll}
    \toprule
    {\bf Experiment}&{\bf Aggreg. data}&{\bf worker/peer data}&{\bf Model param.}&{\bf Forward FLOPs}&{\bf Backward FLOPs}\\
    \midrule
    MLP - master-worker & 10K images & 7.5K images & 52.6K & 105.1K (1 image) & 109.8K (1 image) \\
    MLP - peer to peer & --- & 7.5K train + 10K test images & 52.6K & 105.1K (1 image) & 109.8K (1 image) \\
    YOLO v5n - EI tree & None & 148 frames & 
    1.87M &
    2.68G (1 frame) & --- \\
    \bottomrule
        \end{tabular}
    }       
\end{table*}

\subsubsection{Achieving fairness of comparison}
\new{Many steps have been taken to ensure a fair comparison between the different architectures, since they have different degrees of computational power and maturity: x86-64 and ARM-v8 platforms are server-grade machines, while RISC-V is still a young and experimental embedded-like platform (even if some more HPC-oriented prototypes are being actively developed, such as the Ventana processors).}
First of all, the number of cores available to each machine has been taken into account; since the least powerful platform from this point of view is SiFive RISC-V (4 cores per node), we capped all the processes in the Intel and Ampere experiments consequently, so that precisely 4 cores would be assigned to each of them: this is done through the \texttt{taskset} command.

The maximum number of computational nodes available on the Monte Cimone computational system was 8, so we calibrated our experiments on a federation of a maximum of 8 workers. 
Since the number of Intel and Ampere servers is limited to 4, we place two nodes per server when running 8 node configurations. 
In such cases, we take care to place the node threads in different areas of the processors and near the memory banks to limit the interference between them as much as possible.

Due to the invasive PyTorch threading policy, only using the \texttt{taskset} command is insufficient in this scenario. 
Even if the process is restricted to 4 cores, PyTorch still creates a thread pool of as many threads as available cores. 
This behaviour can lead to many threads that, swapping between each other, can spoil the performance of PyTorch and subsequently ruin the fairness of the comparison.
This problem has been resolved by setting \texttt{OMP\_NUM\_THREADS=4}, thus limiting the OpenMP threads created by PyTorch to 4, making it behave on Ampere exactly like on the SiFive platform.
The \texttt{MKL\_NUM\_THREADS=4} has also been set to prevent the MKL library from creating more threads than the assigned cores to obtain the same behaviour on Intel.

Another precaution to exploit the few cores available at maximum is to use the TCP backend of FastFlow instead of the MPI one. 
The motivation behind this is the computational behaviour of the OpenMPI blocking \texttt{receive}. 
When issued, it actuates a busy waiting policy, directly occupying a core at 100\% of its capability, which would skew the energy results. 
Moreover, in the context of SiFive RISC-V, this means wasting $ 1/4 $ of the available computational power. 
We thus resorted to the TCP backend.

Lastly, a shared workload for the different experiments has to be set. 
Table~\ref{tab:setting} summarises the details of the data used and estimated model complexity.
Due to the massive difference in the computational performance of the various machines, the choice has been made with particular attention to the SiFive RISC-V.
We have chosen to train the MLP on MNIST for 100 epochs, subdivided into 20 federated rounds composed of 5 epochs each. This choice allows the learning curve to stabilise. At the same time, it allows the assessment of relevant measurements (e.g., communication and computation costs). 
\new{We also experimented with a larger ML workload: training a ResNet18 network on the CIFAR10 dataset. Unfortunately, such a configuration required ~24 hours on the RISC-V processor to complete a single training epoch, which is a time span incompatible with our experimental setting.}

A brief video of 148 frames containing people moving has been chosen for the YOLOv5 experiments. We report the mean experimental results across five different runs in Table \ref{tab:results}.
\new{Note that these choices, especially the choice of using only a subset of the available cores, do not hinder x86-64 and ARM platforms in favour of RISC-V. In fact, due to the chosen workload, the total computation time increases with the number of involved cores. This is due to the reduced benefits that are gained in parallelizing the training of a small model and the additional costs required by thread handling and synchronization (in particular under the Python threading model).}



\subsection{PyTorch on RISC-V}
Before this work, the PyTorch framework \cite{PyTorch:19} could not be compiled for the \texttt{riscv} and \texttt{riscv64} CPU architectures. 
In particular, at least three internal dependencies had not yet been ported to the RISC-V ISA: the Chromium Breakpad library\footnote{\url{https://chromium.googlesource.com/breakpad}}, the SLEEF Vectorized Math Library\footnote{\url{https://sleef.org}}, and the PyTorch CPU INFOrmation library (cpuinfo)\footnote{\url{https://github.com/pytorch/cpuinfo}}. 
This work introduces the RISC-V porting of the first two libraries as a byproduct. 
The experiments described in this section also served as functional tests to assess the maturity of the porting. 
In particular, the lack of cpuinfo porting affects some low-level multithreading functions (e.g., \texttt{torch::set\_num\_threads}), and the lack of vector registries in the SiFive Freedom U740 chip affects the SLEEF performance, with Fused Multiply-Add (FMA) as the only optimized instruction for tensor math.

Despite such limitations, the current PyTorch porting is mature enough for research and development. 
On the other hand, efforts to provide a full-featured RISC-V implementation are ongoing. 
The Breakpad porting has already been merged into the official codebase and is publicly available. 
A porting of cpuinfo and a RISC-V vector implementation of the SLEEF library are in the plan for the \db{EuroHPC EUPilot} project. 

\subsection{Results analysis}
The results of the experiments are reported extensively in Table~\ref{tab:results}. 
As can be seen, SiFive is an order of magnitude slower than the other systems, being almost always between 25-35 times slower than Intel and Ampere.
This is to be expected due to the young stage of development of the platform, the absence of optimisaed libraries for deep learning-specific computations, and the lack of vectorial accelerators. 
\new{We believe the lacking of vectorial units to be particularly detrimental to the SiFive processor. The code running on the other two platforms is compiled with libraries explicitly  optimised for exploiting vectorial units (the oneMKL library for Intel and the Arm Performance Library for Ampere).}
On the other hand, the gap in performance between Intel and Ampere is negligible: Ampere is almost as fast as Intel in the computation while consuming an order of magnitude less power. 
The difference in power consumption is discussed more in detail in Section~\ref{sec:power}. 

From a scaling perspective, all experiments behaved as expected: the recorded execution times are congruous with the weak scalability law since they remain constant or slightly increase with the number of processes; this confirms the goodness of the software and the capabilities and flexibility of the \ff{} runtime.

The heterogeneous experiments are another strong point in favour of the \ff{} capabilities: thanks to the Cereal serialisation back-end, it is possible to create a federation in which different workers are hosted on different computational systems. 
This feature is not trivial since moving data from one system to another usually implies conversion issues. 
We succeeded in running a distributed master-worker training across all the computational platforms exploited in this study (Intel, Ampere and SiFive), highlighting the flexibility and compatibility features of the proposed software stack. 
This fact is crucial since the federation of different entities cannot come with the requirement that all entities employ the same underlying computational infrastructure.

\new{While not explicitly relevant to this discussion, it is worth reporting that as a sanity check we measured the classification accuracy of the tested systems. All the proposed FL architectures achieve the expected learning performances. Specifically, the MLP model achieved more than 95\% of accuracy in all configurations, reaching up to 97\% in most runs (YOLOv5 performances are not relevant here, since we used pre-trained models).}

Finally, to give a perspective to the presented performance, we have reproduced one of the proposed scenarios (the master-worker structure with 4 workers) with OpenFL, one of the mature FL frameworks from related work.
To further highlight our effort to the RISC-V developers community, we made an effort to port OpenFL to this computational platform. 
This porting required recompiling in ad-hoc ways the software dependencies (ninja, openblas, grpc/grpcio, and crytography).

As reported in table~\ref{tab:results:mnist}, our implementation on the SiFive can complete the training of 100 epochs over the MNIST dataset 
in 673.70 seconds on average (~11 minutes and 7 seconds). Conversely, OpenFL, with the same model, hyper-parameters, data pre-proces\-sing and distribution, achieved an average running time of 2,486 seconds
(~41 minutes and 26.42 seconds).
\new{Additionally, we repeated the same experiment on the x86-64 platform: even in this case we assessed the efficiency of our implementation (23.56 seconds) with respect to OpenFL (59.15 seconds).}
The difference between the two execution times is stunning, and its motivations have already been discussed throughout the entire paper; this is just another example of how a more high-performance-orientated FL framework would benefit the overall FL research environment.

\begin{table*}[t]
    \caption{Computational performance of proposed FL schema. 
    Each result is averaged over 5 runs. The Intel-Ampere experiments have been executed heterogeneously, with half processes allocated to the Intel cluster and the other half to the Ampere cluster.\label{tab:results}}
    \centering
    \begin{subtable}[t]{\textwidth}
        \centering
        \caption{\textbf{MNIST master-worker training results:} These performance metrics have been taken on a set of 20 federation rounds made up of 5 training epochs each (total 100 epochs); each client was assigned 1/8 of the entire dataset.\label{tab:results:mnist}}
    {\small
            \begin{tabular}{l@{\hspace{12pt}}r
            >{\raggedleft\arraybackslash}p{2.9cm}
            c@{\hspace{12pt}}r
            >{\raggedleft\arraybackslash}p{2.9cm}
            c@{\hspace{12pt}}r
            >{\raggedleft\arraybackslash}p{2.9cm}}
            \toprule
            &\multicolumn{2}{c}{\bf master + 2 workers} && \multicolumn{2}{c}{\bf master + 4 workers}  && \multicolumn{2}{c}{\bf master + 7 workers} \\
            \cmidrule{2-3} \cmidrule{5-6} \cmidrule{8-9}
            & time (s) & energy/worker (J): $\Delta$ (tot) && time (s) & energy/worker (J): $\Delta$ (tot) && Time (s) & energy/worker (J): $\Delta$ (tot) \\
            \midrule
            x86-64 (Intel) & 23.84 & 973 (1992) && \textbf{23.56} & 1011 (2069) && \textbf{24.38} & 1049 (2146) \\ 
            ARM-v8 (Ampere) & \textbf{23.33} & \textbf{133} (\textbf{483}) && 25.66 & \textbf{146} (\textbf{531}) && 25.86 & \textbf{148} (\textbf{535}) \\ 
            RISC-V (SiFive) & 674.47 & 269 (2562) && 673.70 & 269 (2560) && 687.03 & 274 (2610) \\ 
            Intel-Ampere & 29.50 & --- && 29.55 & --- && 33.34 & --- \\
            \bottomrule
            \end{tabular}
        }
    \end{subtable}
    
    \bigskip
    
    \begin{subtable}[t]{\textwidth}
    \centering
    \caption{\textbf{MNIST peer-to-peer training results:} These performance metrics have been taken on a set of 20 federation rounds made up of 5 training epochs each (total 100 epochs); each client was assigned 1/8 of the entire dataset.\label{tab:results:mnist_p2p}}
    {\small
        \begin{tabular}{l@{\hspace{12pt}}r
        >{\raggedleft\arraybackslash}p{2.9cm}
        c@{\hspace{12pt}}r
        >{\raggedleft\arraybackslash}p{2.9cm}
        c@{\hspace{12pt}}r
        >{\raggedleft\arraybackslash}p{2.9cm}}
        \toprule
        &\multicolumn{2}{c}{\bf 2 peers} && \multicolumn{2}{c}{\bf 4 peers}  && \multicolumn{2}{c}{\bf 8 peers} \\
        \cmidrule{2-3} \cmidrule{5-6} \cmidrule{8-9}
        & time (s) & energy/peer (J): $\Delta$ (tot) && time (s) & energy/peer (J): $\Delta$ (tot) && time (s) & energy/peer (J): $\Delta$ (tot) \\
        \midrule
        x86-64 (Intel) & \textbf{23.15} & 2082 (4261) && \textbf{24.05} & 2162 (4422) && \textbf{24.95} & 2210 (4522) \\ 
        ARM-v8 (Ampere) & 24.39 & \textbf{169} (\textbf{535}) && 24.90 & \textbf{173} (\textbf{546})  && 26.65 & \textbf{185} (\textbf{585})\\ 
        RISC-V (SiFive) & 819.35 & 409 (3195) && 815.55 & 407 (3180) && 933.62 & 466 (3641)\\ 
        Intel-Ampere & 45.20 & --- && 39.13 & --- && 50.88 & --- \\
        \bottomrule
        \end{tabular}
        }
    \end{subtable}

    \bigskip
    \begin{subtable}[t]{\textwidth}
    \centering
    \caption{\textbf{YOLO tree-based inference results:} These performance metrics have been obtained by assigning each leaf a video with 148 frames.\label{tab:results:yolo}}
    {\small
        \begin{tabular}{l@{\hspace{12pt}}r
        >{\raggedleft\arraybackslash}p{2.9cm}
        c@{\hspace{12pt}}r
        >{\raggedleft\arraybackslash}p{2.9cm}
        c@{\hspace{12pt}}r
        >{\raggedleft\arraybackslash}p{2.9cm}}
        \toprule
        &\multicolumn{2}{c}{\bf root + 2 leaves} && \multicolumn{2}{c}{\bf root + 4 leaves}  && \multicolumn{2}{c}{\bf root + 7 leaves} \\
        \cmidrule{2-3} \cmidrule{5-6} \cmidrule{8-9}
        & time (s) & energy/leaf (J): $\Delta$ (tot) && time (s) & energy/leaf (J): $\Delta$ (tot) && time (s) & energy/leaf (J): $\Delta$ (tot) \\
        \midrule
        x86-64 (Intel) & \textbf{19.76} & 1520 (2389) && \textbf{19.38} & 1491 (2343) && \textbf{19.01} & 1462 (2298)\\
        ARM-v8 (Ampere) & 37.16 & \textbf{291} (\textbf{848}) && 39.88 & \textbf{312} (\textbf{910}) && 43.15 & \textbf{338} (\textbf{985})\\ 
        RISC-V (SiFive) & 1201.51 & 841 (4926) && 1205.77 & 844 (4943) && 1212.77 & 848 (4972) \\
        Intel-Ampere & 35.65 & --- && 35.65 & --- && 36.10 & --- \\
        \bottomrule
        \end{tabular}
        }
    \end{subtable}
\end{table*}


\begin{table}[t]
    \caption{Comparison of the different systems based on CPU power consumption. These results have been calculated as the mean across the three different configurations of the master-worker scenario.    \label{tab:power}} 
    \centering
    {\small
    \begin{tabular}{l@{}c@{\;\;}c@{\;\;}c@{\;\;}c}
    \toprule
     & \makecell{\bf $\boldmath \Delta$ energy/FLOP\\ \bf (CPU only)} 
     & \makecell{\bf energy \\ \bf /FLOP} 
     & \makecell{\bf avg CPU \\ \bf power (idle)} 
     & \makecell{\bf TDP \\ \bf /socket}\\
    \midrule 
         x86-64 (Intel) & 6.3 nJ & 12.8 nJ & 44 W & 125 W\\
         ARM-v8 (Ampere) & 0.9 nJ & 3.2 nJ & 15 W & 250 W\\
         RISC-V (SiFive) & 1.7 nJ & 15.9 nJ & 3.4 W & 5 W\\
    \bottomrule
    \end{tabular}
    }
\end{table}

\subsection{Power consumption analysis}
\label{sec:power}


Looking at Table~\ref{tab:power}, we can highlight that the three tested systems belong to different computing classes. 
Indeed the table reports idle CPU and system power consumption as well as the Thermal Design Power (TDP) per node and system type. 
The Ampere and Intel CPUs are server processors with a TDP of over 100 Watts, while the SiFive CPU is an embedded class processor with TDP below 5 Watts. 
Table~\ref{tab:results} reports the dynamic energy ($\Delta$) as well as total energy per worker/peer/leaf when training the MNIST model (a,b) and inference of the YOLOv5 model (c).
The Intel and Ampere CPUs have comparable performance, but the Ampere CPU has a lower power consumption of almost an order of magnitude. 
This fact makes the Ampere CPU the most performing one and the most energy-efficient one in this study. 
When compared to the Intel CPU, the Ampere one attains an average reduction for delta and total energy of 8.3x (4.3x) for the MNIST master-worker case, 5.7x (3.7x) for the MNIST peer-to-peer case and 4.7x (2.6x) for the YOLOv5 tree-based case. 
The SiFive CPU, when compared to the Ampere one, attains an average delta (total) energy consumption which is only 1.9x (5.0x) higher for the MNIST master-worker case, 2.4x (6.0x) for the MNIST peer-to-peer case and 2.7x (5.4x) for the YOLOv5 tree-based case. 

\begin{figure}[t]
    \centering
    \includegraphics[width=0.8  \columnwidth]{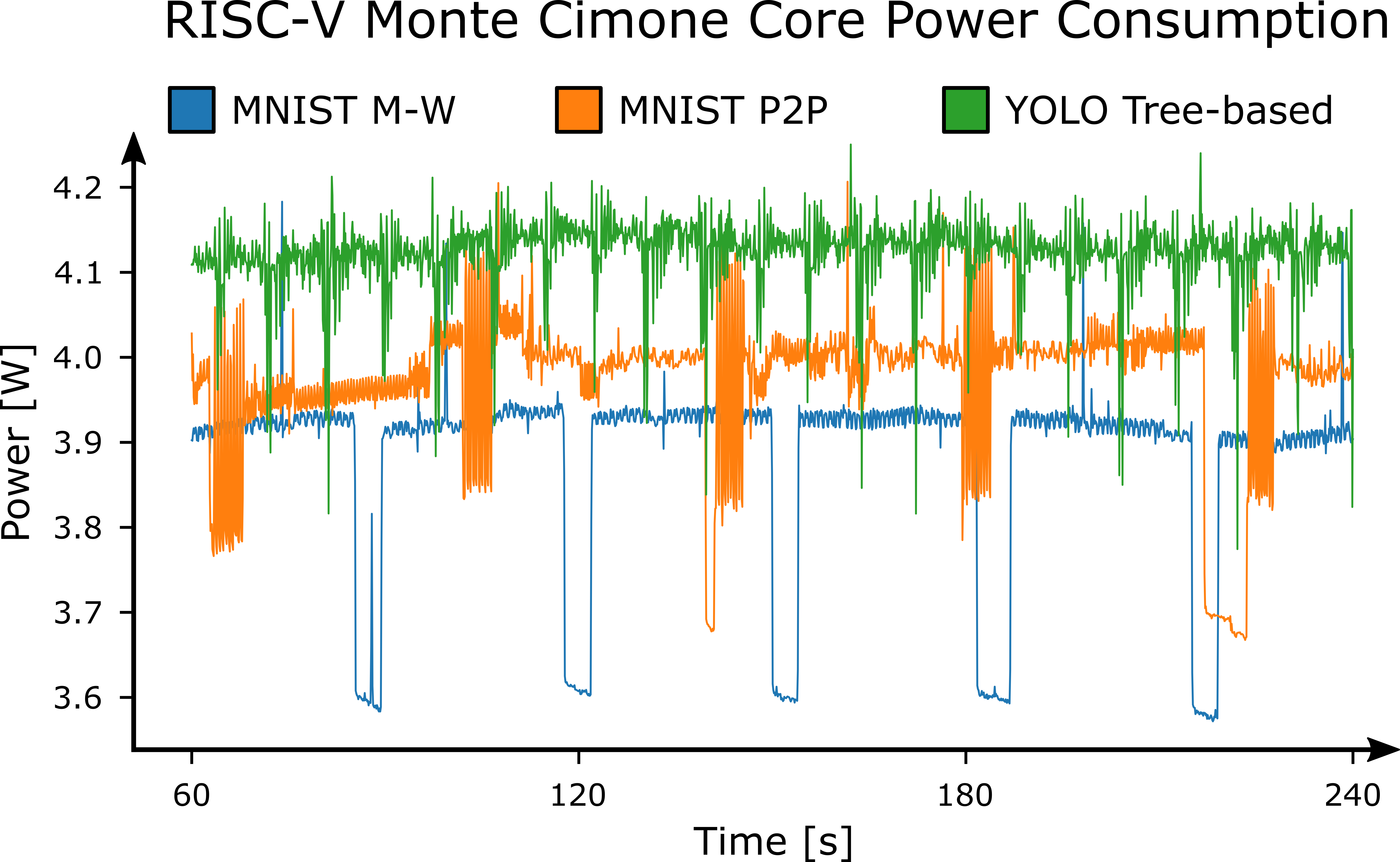}
    \caption{Three minutes of power consumption traces on the Monte Cimone RISC-V system.}
    \label{fig:PowerRISCV}
\end{figure}

While this is an unexpected result, given the absence of SIMD/Vector extension for the SiFive compute nodes, the significantly longer compute times drive up the consumed energy. We expect improved results once new silicon implements the RISC-V ISA vector extensions.
Figure~\ref{fig:PowerRISCV} reports an extract of three minutes of the CPU power consumption (in the y-axis) for a single SiFive compute node for the three experimental settings. 
We can notice that the MNIST master-worker (MNIST M-W in the figure) achieves a significantly lower power consumption than the MNIST peer-to-peer (MNIST P2P in the figure) while having a lower time-to-solution. 
This fact can be explained by the additional computation due to each peer computing its own global model and the increased traffic to handle peer-to-peer communication. 
In both traces, data transfers are visible as lower power consumption segments, each corresponding to a federated round. 
The YOLOv5 tree-based experiment has a higher computational intensity in line with the higher model complexity, which translates into a higher peak power consumption.

When comparing the different configurations, we can notice an overall increase in the time-to-solution while increasing the number of workers/peers/leaves, which directly translates into an increase in the energy consumption of the models. 
The scalability is different for the peer-to-peer and master-worker communication topologies, where the master-worker time-to-solution overhead saturates with the 4 workers case; in contrast, the peer-to-peer time-to-solution overhead continues to scale with the number of peers. 
This behaviour is expected due to the exponential cost in the communication scalability in the latter. 

In Table~\ref{tab:power}, we computed the energy efficiency (Joules per Floating point operation) of the different processors by extracting the floating-point operations for each training epoch and the energy consumed by a single worker in the master-worker scenarios. 
The reported values are calculated following the equation: $ E_{\text{FLOP}} = {P \cdot t_{\text{epoch}}}/[{N_{\text{images}} \cdot (N_{\text{Forward}} + N_{\text{Backward}})]}$ where $P$ is the mean consumed power, $t_{\text{epoch}}$ the time to train one epoch, $N_{\text{images}}$ the number of images, and $N_{\text{Forward}}$, $N_{\text{Backward}}$ the profiled FLOPS for forward and backward pass.
Taking into account only the energy effectively consumed by the calculation itself ($\Delta$ energy), both Ampere and SiFive result to be more energy efficient than the Intel CPU, respectively 7x and 3.7x; this is good, especially for the SiFive platform, given the system's novelty.
On the other hand, if the whole system consumption is taken into account, then the results are dramatically different: in this case, Ampere and Intel result more energy efficient than SiFive, by 5x and 1.2x respectively; this is due to the way higher execution time required by the SiFive platform to complete the same amount of computation done by the other two processors.

\section{Conclusions}

While FL is becoming a popular research topic for all the aspects related to model quality and robustness against model inversion, 
the research on enabling methodologies and architectures for developing new FL systems is lagging. 
Most implementations deeply encode the cooperation protocol and the message serialisation in the framework, making experimentation with new topologies, protocols, and FL schema too complex. 
This work addresses this problem, proposing a lightweight methodology to experiment with FL at the edge, obtained by combining RISC-pb$^2$l and \ff{}. 
We experimented with three different edge ML architectures that can be compiled as shared-memory (for simulation) or distributed-memory (for deployment) versions with the same code 
compatible and interoperable with ARM-v8, RISC-V and x86-64 platforms. 
Notably, as a byproduct, we developed the first working binary of PyTorch for the RISC-V system. 
The experiments' results depict the three processors' maturity levels. 
The SiFive multi-core (4 x U74 RV64GCB) is the most performing RISC-V we found on the market, and it is still far away from mainstream processors, at least on ML workloads. 
Many novel RISC-V implementations, including RISC-V accelerators (e.g., for vectors), are expected to reduce the gap, provided that properly optimised middleware will be available for the whole computing spectrum, including distributed computing and ML. 
\new{This work aims to contribute to the emerging RISC-V ecosystem directly, benchmarking the performance of a current implementation of this ISA in perspective of future high-performance oriented developments, and also porting widely used software to this environment.}


\begin{acks}
This work receives EuroHPC-JU funding under grant no. 101034126, with support from the Horizon2020 programme (the European PILOT). This work is also supported by the Spoke "FutureHPC \& BigData” of the ICSC – Centro Nazionale di Ricerca in "High Performance Computing, Big Data and Quantum Computing", funded by European Union – NextGenerationEU.
\end{acks}

\bibliographystyle{ACM-Reference-Format}
\bibliography{Bibliography/biblio.bib}


\begin{thebibliography}{25}


\ifx \showCODEN    \undefined \def \showCODEN     #1{\unskip}     \fi
\ifx \showDOI      \undefined \def \showDOI       #1{#1}\fi
\ifx \showISBNx    \undefined \def \showISBNx     #1{\unskip}     \fi
\ifx \showISBNxiii \undefined \def \showISBNxiii  #1{\unskip}     \fi
\ifx \showISSN     \undefined \def \showISSN      #1{\unskip}     \fi
\ifx \showLCCN     \undefined \def \showLCCN      #1{\unskip}     \fi
\ifx \shownote     \undefined \def \shownote      #1{#1}          \fi
\ifx \showarticletitle \undefined \def \showarticletitle #1{#1}   \fi
\ifx \showURL      \undefined \def \showURL       {\relax}        \fi
\providecommand\bibfield[2]{#2}
\providecommand\bibinfo[2]{#2}
\providecommand\natexlab[1]{#1}
\providecommand\showeprint[2][]{arXiv:#2}

\bibitem[\protect\citeauthoryear{Adolf, Rama, Reagen, Wei, and Brooks}{Adolf
  et~al\mbox{.}}{2016}]%
        {adolf16fathom}
\bibfield{author}{\bibinfo{person}{Robert Adolf}, \bibinfo{person}{Saketh
  Rama}, \bibinfo{person}{Brandon Reagen}, \bibinfo{person}{Gu-Yeon Wei}, {and}
  \bibinfo{person}{David Brooks}.} \bibinfo{year}{2016}\natexlab{}.
\newblock \showarticletitle{Fathom: Reference workloads for modern deep
  learning methods}. In \bibinfo{booktitle}{\emph{2016 IEEE International
  Symposium on Workload Characterization (IISWC)}}. \bibinfo{publisher}{IEEE},
  \bibinfo{address}{Providence, RI, USA}, \bibinfo{pages}{1--10}.
\newblock
\showISSN{978-1-5090-3897-8}
\urldef\tempurl%
\url{https://doi.org/10.1109/IISWC.2016.7581275}
\showDOI{\tempurl}


\bibitem[\protect\citeauthoryear{Aldinucci, Campa, Danelutto, Kilpatrick, and
  Torquati}{Aldinucci et~al\mbox{.}}{2013}]%
        {aldinucci2014design}
\bibfield{author}{\bibinfo{person}{Marco Aldinucci}, \bibinfo{person}{Sonia
  Campa}, \bibinfo{person}{Marco Danelutto}, \bibinfo{person}{Peter
  Kilpatrick}, {and} \bibinfo{person}{Massimo Torquati}.}
  \bibinfo{year}{2013}\natexlab{}.
\newblock \showarticletitle{Design patterns percolating to parallel programming
  framework implementation}.
\newblock \bibinfo{journal}{\emph{International Journal of Parallel
  Programming}} \bibinfo{volume}{42}, \bibinfo{number}{6} (\bibinfo{date}{26
  September} \bibinfo{year}{2013}), \bibinfo{pages}{1012--1031}.
\newblock
\urldef\tempurl%
\url{https://doi.org/10.1007/s10766-013-0273-6}
\showDOI{\tempurl}


\bibitem[\protect\citeauthoryear{Aldinucci, Danelutto, Kilpatrick, Meneghin,
  and Torquati}{Aldinucci et~al\mbox{.}}{2012}]%
        {ffqueue-EuroPar2012}
\bibfield{author}{\bibinfo{person}{Marco Aldinucci}, \bibinfo{person}{Marco
  Danelutto}, \bibinfo{person}{Peter Kilpatrick}, \bibinfo{person}{Massimiliano
  Meneghin}, {and} \bibinfo{person}{Massimo Torquati}.}
  \bibinfo{year}{2012}\natexlab{}.
\newblock \showarticletitle{An Efficient Unbounded Lock-Free Queue for
  Multi-core Systems}. In \bibinfo{booktitle}{\emph{Euro-Par 2012 Parallel
  Processing}}, Vol.~\bibinfo{volume}{7484}. \bibinfo{publisher}{Springer
  Berlin Heidelberg}, \bibinfo{address}{Berlin, Heidelberg},
  \bibinfo{pages}{662--673}.
\newblock
\showISBNx{978-3-642-32820-6}
\urldef\tempurl%
\url{https://doi.org/10.1007/978-3-642-32820-6_65}
\showDOI{\tempurl}


\bibitem[\protect\citeauthoryear{Aldinucci, Danelutto, Kilpatrick, and
  Torquati}{Aldinucci et~al\mbox{.}}{2017}]%
        {aldinucci2017fastflow}
\bibfield{author}{\bibinfo{person}{Marco Aldinucci}, \bibinfo{person}{Marco
  Danelutto}, \bibinfo{person}{Peter Kilpatrick}, {and}
  \bibinfo{person}{Massimo Torquati}.} \bibinfo{year}{2017}\natexlab{}.
\newblock \bibinfo{booktitle}{\emph{Fastflow: High-Level and Efficient
  Streaming on Multicore}}.
\newblock \bibinfo{publisher}{Wiley-Blackwell}, \bibinfo{pages}{261--280}.
\newblock


\bibitem[\protect\citeauthoryear{Authors}{Authors}{2019}]%
        {tff}
\bibfield{author}{\bibinfo{person}{The TensorFlow~Federated Authors}.}
  \bibinfo{year}{2019}\natexlab{}.
\newblock \bibinfo{booktitle}{\emph{TensorFlow Federated}}.
\newblock Google.
\newblock
\urldef\tempurl%
\url{https://github.com/tensorflow/federated}
\showURL{%
\tempurl}


\bibitem[\protect\citeauthoryear{Bartolini, Ficarelli, Parisi, Beneventi,
  Barchi, Gregori, Magugliani, Cicala, Gianfreda, Cesarini,
  et~al\mbox{.}}{Bartolini et~al\mbox{.}}{2022}]%
        {mcimone}
\bibfield{author}{\bibinfo{person}{Andrea Bartolini}, \bibinfo{person}{Federico
  Ficarelli}, \bibinfo{person}{Emanuele Parisi}, \bibinfo{person}{Francesco
  Beneventi}, \bibinfo{person}{Francesco Barchi}, \bibinfo{person}{Daniele
  Gregori}, \bibinfo{person}{Fabrizio Magugliani}, \bibinfo{person}{Marco
  Cicala}, \bibinfo{person}{Cosimo Gianfreda}, \bibinfo{person}{Daniele
  Cesarini}, {et~al\mbox{.}}} \bibinfo{year}{2022}\natexlab{}.
\newblock \showarticletitle{Monte Cimone: Paving the Road for the First
  Generation of RISC-V High-Performance Computers}. In
  \bibinfo{booktitle}{\emph{2022 IEEE 35th International System-on-Chip
  Conference (SOCC)}}. \bibinfo{publisher}{IEEE}, \bibinfo{address}{Belfast,
  United Kingdom}, \bibinfo{pages}{1--6}.
\newblock
\showISSN{2164-1706}
\urldef\tempurl%
\url{https://doi.org/10.1109/SOCC56010.2022.9908096}
\showDOI{\tempurl}


\bibitem[\protect\citeauthoryear{Beltr{\'a}n, P{\'e}rez, S{\'a}nchez, Bernal,
  Bovet, P{\'e}rez, P{\'e}rez, and Celdr{\'a}n}{Beltr{\'a}n
  et~al\mbox{.}}{2022}]%
        {beltran2022decentralized}
\bibfield{author}{\bibinfo{person}{Enrique Tom{\'a}s~Mart{\'\i}nez
  Beltr{\'a}n}, \bibinfo{person}{Mario~Quiles P{\'e}rez},
  \bibinfo{person}{Pedro Miguel~S{\'a}nchez S{\'a}nchez},
  \bibinfo{person}{Sergio~L{\'o}pez Bernal}, \bibinfo{person}{G{\'e}r{\^o}me
  Bovet}, \bibinfo{person}{Manuel~Gil P{\'e}rez},
  \bibinfo{person}{Gregorio~Mart{\'\i}nez P{\'e}rez}, {and}
  \bibinfo{person}{Alberto~Huertas Celdr{\'a}n}.}
  \bibinfo{year}{2022}\natexlab{}.
\newblock \showarticletitle{Decentralized Federated Learning: Fundamentals,
  State-of-the-art, Frameworks, Trends, and Challenges}.
\newblock \bibinfo{journal}{\emph{arXiv preprint arXiv:2211.08413}}
  (\bibinfo{date}{15 November} \bibinfo{year}{2022}).
\newblock
\urldef\tempurl%
\url{https://doi.org/10.48550/ARXIV.2211.08413}
\showDOI{\tempurl}


\bibitem[\protect\citeauthoryear{Beutel, Topal, Mathur, Qiu, Parcollet,
  de~Gusm{\~a}o, and Lane}{Beutel et~al\mbox{.}}{2020}]%
        {beutel2020flower}
\bibfield{author}{\bibinfo{person}{Daniel~J Beutel}, \bibinfo{person}{Taner
  Topal}, \bibinfo{person}{Akhil Mathur}, \bibinfo{person}{Xinchi Qiu},
  \bibinfo{person}{Titouan Parcollet}, \bibinfo{person}{Pedro~PB de
  Gusm{\~a}o}, {and} \bibinfo{person}{Nicholas~D Lane}.}
  \bibinfo{year}{2020}\natexlab{}.
\newblock \showarticletitle{Flower: A friendly federated learning research
  framework}.
\newblock \bibinfo{journal}{\emph{arXiv preprint arXiv:2007.14390}}
  (\bibinfo{date}{28 July} \bibinfo{year}{2020}).
\newblock
\urldef\tempurl%
\url{https://doi.org/10.48550/ARXIV.2007.14390}
\showDOI{\tempurl}


\bibitem[\protect\citeauthoryear{Caldas, Duddu, Wu, Li, Kone{\v{c}}n{\`y},
  McMahan, Smith, and Talwalkar}{Caldas et~al\mbox{.}}{2018}]%
        {caldas2018leaf}
\bibfield{author}{\bibinfo{person}{Sebastian Caldas}, \bibinfo{person}{Sai
  Meher~Karthik Duddu}, \bibinfo{person}{Peter Wu}, \bibinfo{person}{Tian Li},
  \bibinfo{person}{Jakub Kone{\v{c}}n{\`y}}, \bibinfo{person}{H~Brendan
  McMahan}, \bibinfo{person}{Virginia Smith}, {and} \bibinfo{person}{Ameet
  Talwalkar}.} \bibinfo{year}{2018}\natexlab{}.
\newblock \showarticletitle{Leaf: A benchmark for federated settings}.
\newblock \bibinfo{journal}{\emph{arXiv preprint arXiv:1812.01097}}
  (\bibinfo{date}{3 December} \bibinfo{year}{2018}).
\newblock
\urldef\tempurl%
\url{https://doi.org/10.48550/ARXIV.1812.01097}
\showDOI{\tempurl}


\bibitem[\protect\citeauthoryear{Chen and Sun}{Chen and Sun}{2021}]%
        {chen2021understanding}
\bibfield{author}{\bibinfo{person}{Jihong Chen} {and} \bibinfo{person}{Jiabin
  Sun}.} \bibinfo{year}{2021}\natexlab{}.
\newblock \showarticletitle{Understanding the Chinese Data Security Law}.
\newblock \bibinfo{journal}{\emph{International Cybersecurity Law Review}}
  \bibinfo{volume}{2}, \bibinfo{number}{2} (\bibinfo{date}{12 October}
  \bibinfo{year}{2021}), \bibinfo{pages}{209--221}.
\newblock
\urldef\tempurl%
\url{https://doi.org/10.1365/s43439-021-00038-3}
\showDOI{\tempurl}


\bibitem[\protect\citeauthoryear{Cheng, Fan, Jin, Liu, Chen, Papadopoulos, and
  Yang}{Cheng et~al\mbox{.}}{2021}]%
        {cheng2021secureboost}
\bibfield{author}{\bibinfo{person}{Kewei Cheng}, \bibinfo{person}{Tao Fan},
  \bibinfo{person}{Yilun Jin}, \bibinfo{person}{Yang Liu},
  \bibinfo{person}{Tianjian Chen}, \bibinfo{person}{Dimitrios Papadopoulos},
  {and} \bibinfo{person}{Qiang Yang}.} \bibinfo{year}{2021}\natexlab{}.
\newblock \showarticletitle{Secureboost: A lossless federated learning
  framework}.
\newblock \bibinfo{journal}{\emph{IEEE Intelligent Systems}}
  \bibinfo{volume}{36}, \bibinfo{number}{6} (\bibinfo{date}{25 May}
  \bibinfo{year}{2021}), \bibinfo{pages}{87--98}.
\newblock
\showISSN{1941-1294}
\urldef\tempurl%
\url{https://doi.org/10.1109/MIS.2021.3082561}
\showDOI{\tempurl}


\bibitem[\protect\citeauthoryear{Goodman and Flaxman}{Goodman and
  Flaxman}{2017}]%
        {goodman2017european}
\bibfield{author}{\bibinfo{person}{Bryce Goodman} {and} \bibinfo{person}{Seth
  Flaxman}.} \bibinfo{year}{2017}\natexlab{}.
\newblock \showarticletitle{European Union Regulations on Algorithmic
  Decision-Making and a "Right to Explanation"}.
\newblock \bibinfo{journal}{\emph{AI Magazine}} \bibinfo{volume}{38},
  \bibinfo{number}{3} (\bibinfo{date}{2 Oct} \bibinfo{year}{2017}),
  \bibinfo{pages}{50--57}.
\newblock
\showISSN{0738-4602}
\urldef\tempurl%
\url{https://doi.org/10.1609/aimag.v38i3.2741}
\showDOI{\tempurl}


\bibitem[\protect\citeauthoryear{Grant and Voorhies}{Grant and
  Voorhies}{2013}]%
        {grant2013cereal}
\bibfield{author}{\bibinfo{person}{Shane~W. Grant} {and}
  \bibinfo{person}{Randolph Voorhies}.} \bibinfo{year}{2013}\natexlab{}.
\newblock \bibinfo{booktitle}{\emph{Cereal a C++11 library for serialization}}.
\newblock USCiLab.
\newblock
\urldef\tempurl%
\url{https://github.com/USCiLab/cereal}
\showURL{%
\tempurl}


\bibitem[\protect\citeauthoryear{He, Li, So, Zeng, Zhang, Wang, Wang,
  Vepakomma, Singh, Qiu, et~al\mbox{.}}{He et~al\mbox{.}}{2020}]%
        {he2020fedml}
\bibfield{author}{\bibinfo{person}{Chaoyang He}, \bibinfo{person}{Songze Li},
  \bibinfo{person}{Jinhyun So}, \bibinfo{person}{Xiao Zeng},
  \bibinfo{person}{Mi Zhang}, \bibinfo{person}{Hongyi Wang},
  \bibinfo{person}{Xiaoyang Wang}, \bibinfo{person}{Praneeth Vepakomma},
  \bibinfo{person}{Abhishek Singh}, \bibinfo{person}{Hang Qiu},
  {et~al\mbox{.}}} \bibinfo{year}{2020}\natexlab{}.
\newblock \showarticletitle{Fedml: A research library and benchmark for
  federated machine learning}.
\newblock \bibinfo{journal}{\emph{arXiv preprint arXiv:2007.13518}}
  (\bibinfo{date}{27 July} \bibinfo{year}{2020}).
\newblock
\urldef\tempurl%
\url{https://doi.org/10.48550/ARXIV.2007.13518}
\showDOI{\tempurl}


\bibitem[\protect\citeauthoryear{Li, Fan, Tse, and Lin}{Li
  et~al\mbox{.}}{2020a}]%
        {li2020review}
\bibfield{author}{\bibinfo{person}{Li Li}, \bibinfo{person}{Yuxi Fan},
  \bibinfo{person}{Mike Tse}, {and} \bibinfo{person}{Kuo-Yi Lin}.}
  \bibinfo{year}{2020}\natexlab{a}.
\newblock \showarticletitle{A review of applications in federated learning}.
\newblock \bibinfo{journal}{\emph{Computers \& Industrial Engineering}}
  \bibinfo{volume}{149} (\bibinfo{date}{November} \bibinfo{year}{2020}),
  \bibinfo{pages}{106854}.
\newblock
\showISSN{0360-8352}
\urldef\tempurl%
\url{https://doi.org/10.1016/j.cie.2020.106854}
\showDOI{\tempurl}


\bibitem[\protect\citeauthoryear{Li, Zhao, Varma, Salpekar, Noordhuis, Li,
  Paszke, Smith, Vaughan, Damania, and Chintala}{Li et~al\mbox{.}}{2020b}]%
        {li2020pytorch}
\bibfield{author}{\bibinfo{person}{Shen Li}, \bibinfo{person}{Yanli Zhao},
  \bibinfo{person}{Rohan Varma}, \bibinfo{person}{Omkar Salpekar},
  \bibinfo{person}{Pieter Noordhuis}, \bibinfo{person}{Teng Li},
  \bibinfo{person}{Adam Paszke}, \bibinfo{person}{Jeff Smith},
  \bibinfo{person}{Brian Vaughan}, \bibinfo{person}{Pritam Damania}, {and}
  \bibinfo{person}{Soumith Chintala}.} \bibinfo{year}{2020}\natexlab{b}.
\newblock \showarticletitle{Pytorch distributed: Experiences on accelerating
  data parallel training}.
\newblock \bibinfo{journal}{\emph{arXiv preprint arXiv:2006.15704}}
  (\bibinfo{date}{28 June} \bibinfo{year}{2020}).
\newblock
\urldef\tempurl%
\url{https://doi.org/10.48550/ARXIV.2006.15704}
\showDOI{\tempurl}


\bibitem[\protect\citeauthoryear{McMahan, Moore, Ramage, Hampson, and
  Ag{\"{u}}era~y Arcas}{McMahan et~al\mbox{.}}{2017}]%
        {mcmahan2017communication}
\bibfield{author}{\bibinfo{person}{Brendan McMahan}, \bibinfo{person}{Eider
  Moore}, \bibinfo{person}{Daniel Ramage}, \bibinfo{person}{Seth Hampson},
  {and} \bibinfo{person}{Blaise Ag{\"{u}}era~y Arcas}.}
  \bibinfo{year}{2017}\natexlab{}.
\newblock \showarticletitle{Communication-Efficient Learning of Deep Networks
  from Decentralized Data}. In \bibinfo{booktitle}{\emph{Proceedings of the
  20th International Conference on Artificial Intelligence and Statistics
  {AISTATS}}} \emph{(\bibinfo{series}{Proceedings of Machine Learning
  Research}, Vol.~\bibinfo{volume}{54})},
  \bibfield{editor}{\bibinfo{person}{Aarti Singh} {and}
  \bibinfo{person}{Xiaojin~(Jerry) Zhu}} (Eds.). \bibinfo{publisher}{PMLR},
  \bibinfo{address}{Fort Lauderdale, FL, USA}, \bibinfo{pages}{1273--1282}.
\newblock
\showISBNx{9781713807933}


\bibitem[\protect\citeauthoryear{Pardau}{Pardau}{2018}]%
        {pardau2018california}
\bibfield{author}{\bibinfo{person}{Stuart~L Pardau}.}
  \bibinfo{year}{2018}\natexlab{}.
\newblock \showarticletitle{The California consumer privacy act: towards a
  European-style privacy regime in the United States}.
\newblock \bibinfo{journal}{\emph{J. Tech. L. \& Pol'y}} \bibinfo{volume}{23},
  \bibinfo{number}{1} (\bibinfo{year}{2018}), \bibinfo{pages}{68--114}.
\newblock
\showISSN{1087-6995}


\bibitem[\protect\citeauthoryear{Paszke, Gross, Massa, Lerer, Bradbury, Chanan,
  Killeen, Lin, Gimelshein, Antiga, Desmaison, K{\"{o}}pf, Yang, DeVito,
  Raison, Tejani, Chilamkurthy, Steiner, Fang, Bai, and Chintala}{Paszke
  et~al\mbox{.}}{2019}]%
        {PyTorch:19}
\bibfield{author}{\bibinfo{person}{Adam Paszke}, \bibinfo{person}{Sam Gross},
  \bibinfo{person}{Francisco Massa}, \bibinfo{person}{Adam Lerer},
  \bibinfo{person}{James Bradbury}, \bibinfo{person}{Gregory Chanan},
  \bibinfo{person}{Trevor Killeen}, \bibinfo{person}{Zeming Lin},
  \bibinfo{person}{Natalia Gimelshein}, \bibinfo{person}{Luca Antiga},
  \bibinfo{person}{Alban Desmaison}, \bibinfo{person}{Andreas K{\"{o}}pf},
  \bibinfo{person}{Edward~Z. Yang}, \bibinfo{person}{Zachary DeVito},
  \bibinfo{person}{Martin Raison}, \bibinfo{person}{Alykhan Tejani},
  \bibinfo{person}{Sasank Chilamkurthy}, \bibinfo{person}{Benoit Steiner},
  \bibinfo{person}{Lu Fang}, \bibinfo{person}{Junjie Bai}, {and}
  \bibinfo{person}{Soumith Chintala}.} \bibinfo{year}{2019}\natexlab{}.
\newblock \showarticletitle{PyTorch: An Imperative Style, High-Performance Deep
  Learning Library}. In \bibinfo{booktitle}{\emph{Advances in Neural
  Information Processing Systems 32: Annual Conference on Neural Information
  Processing Systems 2019, NeurIPS 2019}},
  \bibfield{editor}{\bibinfo{person}{Hanna~M. Wallach}, \bibinfo{person}{Hugo
  Larochelle}, \bibinfo{person}{Alina Beygelzimer}, \bibinfo{person}{Florence
  d'Alch{\'{e}}{-}Buc}, \bibinfo{person}{Emily~B. Fox}, {and}
  \bibinfo{person}{Roman Garnett}} (Eds.), Vol.~\bibinfo{volume}{32}.
  \bibinfo{address}{Vancouver Convention Center, Vancouver CANADA},
  \bibinfo{pages}{8024--8035}.
\newblock
\showISBNx{9781713807933}


\bibitem[\protect\citeauthoryear{Reina, Gruzdev, Foley, Perepelkina, Sharma,
  Davidyuk, Trushkin, Radionov, Mokrov, Agapov, Martin, Edwards, Sheller, Pati,
  Moorthy, Wang, Shah, and Bakas}{Reina et~al\mbox{.}}{2021}]%
        {reina2021openfl}
\bibfield{author}{\bibinfo{person}{G.~Anthony Reina}, \bibinfo{person}{Alexey
  Gruzdev}, \bibinfo{person}{Patrick Foley}, \bibinfo{person}{Olga
  Perepelkina}, \bibinfo{person}{Mansi Sharma}, \bibinfo{person}{Igor
  Davidyuk}, \bibinfo{person}{Ilya Trushkin}, \bibinfo{person}{Maksim
  Radionov}, \bibinfo{person}{Aleksandr Mokrov}, \bibinfo{person}{Dmitry
  Agapov}, \bibinfo{person}{Jason Martin}, \bibinfo{person}{Brandon Edwards},
  \bibinfo{person}{Micah~J. Sheller}, \bibinfo{person}{Sarthak Pati},
  \bibinfo{person}{Prakash~Narayana Moorthy}, \bibinfo{person}{Shih-han Wang},
  \bibinfo{person}{Prashant Shah}, {and} \bibinfo{person}{Spyridon Bakas}.}
  \bibinfo{year}{2021}\natexlab{}.
\newblock \showarticletitle{OpenFL: An open-source framework for Federated
  Learning}.
\newblock \bibinfo{journal}{\emph{arXiv preprint arXiv:2105.06413}}
  (\bibinfo{date}{13 May} \bibinfo{year}{2021}).
\newblock
\urldef\tempurl%
\url{https://doi.org/10.48550/arXiv.2105.06413}
\showDOI{\tempurl}


\bibitem[\protect\citeauthoryear{Tonci, Torquati, Mencagli, and
  Danelutto}{Tonci et~al\mbox{.}}{2023}]%
        {Tonci2022DFF}
\bibfield{author}{\bibinfo{person}{Nicol\`{o} Tonci}, \bibinfo{person}{Massimo
  Torquati}, \bibinfo{person}{Gabriele Mencagli}, {and} \bibinfo{person}{Marco
  Danelutto}.} \bibinfo{year}{2023}\natexlab{}.
\newblock \showarticletitle{Distributed-memory FastFlow Building Blocks}.
\newblock \bibinfo{journal}{\emph{International of Parallel Programming}}
  \bibinfo{volume}{51} (\bibinfo{date}{2 December} \bibinfo{year}{2023}),
  \bibinfo{pages}{1--21}.
\newblock
\urldef\tempurl%
\url{https://doi.org/10.1007/s10766-022-00750-5}
\showDOI{\tempurl}


\bibitem[\protect\citeauthoryear{Torquati}{Torquati}{2019}]%
        {TorquatiPhD}
\bibfield{author}{\bibinfo{person}{Massimo Torquati}.}
  \bibinfo{year}{2019}\natexlab{}.
\newblock \emph{\bibinfo{title}{Harnessing Parallelism in Multi/Many-Cores with
  Streams and Parallel Patterns}}.
\newblock \bibinfo{thesistype}{Ph.D. Dissertation}. \bibinfo{school}{University
  of Pisa}.
\newblock


\bibitem[\protect\citeauthoryear{Wu, Brooks, Chen, Chen, Choudhury, Dukhan,
  Hazelwood, Isaac, Jia, Jia, Leyvand, Lu, Lu, Qiao, Reagen, Spisak, Sun,
  Tulloch, Vajda, Wang, Wang, Wasti, Wu, Xian, Yoo, and Zhang}{Wu
  et~al\mbox{.}}{2019}]%
        {wu2019machine}
\bibfield{author}{\bibinfo{person}{Carole-Jean Wu}, \bibinfo{person}{David
  Brooks}, \bibinfo{person}{Kevin Chen}, \bibinfo{person}{Douglas Chen},
  \bibinfo{person}{Sy Choudhury}, \bibinfo{person}{Marat Dukhan},
  \bibinfo{person}{Kim Hazelwood}, \bibinfo{person}{Eldad Isaac},
  \bibinfo{person}{Yangqing Jia}, \bibinfo{person}{Bill Jia},
  \bibinfo{person}{Tommer Leyvand}, \bibinfo{person}{Hao Lu},
  \bibinfo{person}{Yang Lu}, \bibinfo{person}{Lin Qiao},
  \bibinfo{person}{Brandon Reagen}, \bibinfo{person}{Joe Spisak},
  \bibinfo{person}{Fei Sun}, \bibinfo{person}{Andrew Tulloch},
  \bibinfo{person}{Peter Vajda}, \bibinfo{person}{Xiaodong Wang},
  \bibinfo{person}{Yanghan Wang}, \bibinfo{person}{Bram Wasti},
  \bibinfo{person}{Yiming Wu}, \bibinfo{person}{Ran Xian},
  \bibinfo{person}{Sungjoo Yoo}, {and} \bibinfo{person}{Peizhao Zhang}.}
  \bibinfo{year}{2019}\natexlab{}.
\newblock \showarticletitle{Machine learning at facebook: Understanding
  inference at the edge}. In \bibinfo{booktitle}{\emph{2019 IEEE international
  symposium on high performance computer architecture (HPCA)}}. IEEE,
  \bibinfo{address}{Washington, DC, USA}, \bibinfo{pages}{331--344}.
\newblock
\showISSN{2378-203X}
\urldef\tempurl%
\url{https://doi.org/10.1109/HPCA.2019.00048}
\showDOI{\tempurl}


\bibitem[\protect\citeauthoryear{Xie, Wang, Chen, Gao, Yao, Kuang, Li, Ding,
  and Zhou}{Xie et~al\mbox{.}}{2022}]%
        {xie2022federatedscope}
\bibfield{author}{\bibinfo{person}{Yuexiang Xie}, \bibinfo{person}{Zhen Wang},
  \bibinfo{person}{Daoyuan Chen}, \bibinfo{person}{Dawei Gao},
  \bibinfo{person}{Liuyi Yao}, \bibinfo{person}{Weirui Kuang},
  \bibinfo{person}{Yaliang Li}, \bibinfo{person}{Bolin Ding}, {and}
  \bibinfo{person}{Jingren Zhou}.} \bibinfo{year}{2022}\natexlab{}.
\newblock \showarticletitle{Federatedscope: A comprehensive and flexible
  federated learning platform via message passing}.
\newblock \bibinfo{journal}{\emph{arXiv preprint arXiv:2204.05011}}
  (\bibinfo{date}{11 April} \bibinfo{year}{2022}).
\newblock
\urldef\tempurl%
\url{https://doi.org/10.48550/ARXIV.2204.05011}
\showDOI{\tempurl}


\bibitem[\protect\citeauthoryear{Ziller, Trask, Lopardo, Szymkow, Wagner,
  Bluemke, Nounahon, Passerat-Palmbach, Prakash, Rose, et~al\mbox{.}}{Ziller
  et~al\mbox{.}}{2021}]%
        {ziller2021pysyft}
\bibfield{author}{\bibinfo{person}{Alexander Ziller}, \bibinfo{person}{Andrew
  Trask}, \bibinfo{person}{Antonio Lopardo}, \bibinfo{person}{Benjamin
  Szymkow}, \bibinfo{person}{Bobby Wagner}, \bibinfo{person}{Emma Bluemke},
  \bibinfo{person}{Jean-Mickael Nounahon}, \bibinfo{person}{Jonathan
  Passerat-Palmbach}, \bibinfo{person}{Kritika Prakash}, \bibinfo{person}{Nick
  Rose}, {et~al\mbox{.}}} \bibinfo{year}{2021}\natexlab{}.
\newblock \showarticletitle{Pysyft: A library for easy federated learning}.
\newblock \bibinfo{journal}{\emph{Federated Learning Systems: Towards
  Next-Generation AI}}  \bibinfo{volume}{965} (\bibinfo{date}{12 June}
  \bibinfo{year}{2021}), \bibinfo{pages}{111--139}.
\newblock
\showISSN{978-3-030-70604-3}
\urldef\tempurl%
\url{https://doi.org/10.1007/978-3-030-70604-3_5}
\showDOI{\tempurl}


\end{thebibliography}

\clearpage

\appendix
\section{Artifact}
\label{sec:artifact}

\subsection{Abstract}

Fast Federated Learning (FFL) is a C/C++-based Federated Learning framework built on top of the parallel programming FastFlow framework. 
It exploits the Cereal library to efficiently serialise the updates sent over the network and the libtorch library to bypass the need for Python code fully. 
It has been successfully tested on x86\_64, ARM and RISC-V platforms.
FFL has scripts for automatically installing the framework and reproducing all the experiments reported in the original paper.

\subsection{Artifact check-list (meta-information)}

{\em Obligatory. Use just a few informal keywords in all fields applicable to your artifacts
, and remove the rest. This information is needed to find appropriate reviewers and gradually 
unify artifact meta information in Digital Libraries.}

{\small
\begin{itemize}
  \item {\bf Algorithm: }Federated Averaging (FedAvg)
  \item {\bf Compilation: }C++17 compatible compiler
  \item {\bf Binary: } to be compiled from source
  \item {\bf Model: }Multi-Layer Perceptron (MLP), YOLO v5n
  \item {\bf Data set: }MNIST, Ranger Roll (video)
  \item {\bf Run-time environment: }Linux, MacOS, FastFlow, Cereal, OpenCV
  \item {\bf Hardware: }x86\_64, ARM, RISC-V, power consumption counters
  \item {\bf Execution: }sole user
  \item {\bf Metrics: }time, accuracy
  \item {\bf Output: }console: mean execution time, logs: time, accuracy
  \item {\bf Experiments: } README, scripts
  \item {\bf How much disk space required (approximately)?: }874 MB
  \item {\bf How much time is needed to prepare workflow (approximately)?: }10 minutes
  \item {\bf How much time is needed to complete experiments (approximately)?: }~20 minutes on Intel, ~25 minutes and ARM and ~800 minutes on RISC-V
  \item {\bf Publicly available?: }yes
  \item {\bf Code licenses (if publicly available)?: }GPL-3.0
  \item {\bf Data licenses (if publicly available)?: }GPL-3.0, CC BY
  \item {\bf Archived (provide DOI)?: }10.5281/zenodo.7807974
\end{itemize}
}

\subsection{Description}

\subsubsection{How to access}

FFL can be obtained by simply cloning the official GitHub repository: \url{https://github.com/alpha-unito/FastFederatedLearning.git}.
The approximate disk occupation after the setup and compilation is approximately 874 MB.

\subsubsection{Hardware dependencies}

No specific hardware dependency is needed to run FFL.
However, to reproduce the experimental results proposed in the paper, it is necessary to have access to 4 Supermicro servers including 2 Intel Xeon Gold 6230 CPU (20-
cores@2.10GHz) sockets with 1536GB RAM, 4 ARM-dev kits including one socket Ampere-Altra
Q80-30 (80-core@3GHz) with 512GB RAM, and 8 U740 SoC from SiFive running up to 1.2 GHz and 16GB. 
Intel and Ampere CPUs nodes should be interconnected with an Infiniband network, while the SiFive ones with a 1Gb/s Ethernet. 
Also, hardware power monitoring counter should be available for the Intel and ARM platforms, while an external monitoring board, such as the PXIe-4309 National Instruments, is required to record the power consumption on RISC-V.

\subsubsection{Software dependencies}

Starting from a fresh Ubuntu 22.04 installation, the first step to reproduce the reported results reported in the paper is to use \texttt{apt} to update the available package information and install (tested versions are reported in brackets): build-essential (12.9), cmake (3.22.1), libopencv-dev (4.5.4), and unzip (6.0-26) packages.
A C/C++-17 compatible compiler is required for compilation.
Furthermore, the FastFlow (DistributedFF branch), Cereal (1.3.2), and libtorch (2.0.0) libraries are required (installed via helper script see \S\ref{ssec:installation}).
Optionally, the Powermon utility is required to measure system power consumption.

\subsubsection{Data sets}

The MNIST dataset and a short video (Ranger Roll) are needed for the experiments.
The setup script automatically downloads these files: MNIST is retrieved from the owner's website, while the Ranger Roll video is downloaded from our servers.

\subsubsection{Models}

A simple, three-layer MLP is used for part of the experiments, while the others exploit a YOLO v5n neural network. Our artifact provides both models.

\subsection{Installation}
\label{ssec:installation}
To set up the whole system and to compile the example, it is sufficient to run: \texttt{source setup.sh}.	
This script will automatically download all the required libraries, update the environment variables, build the \texttt{dff\_run} utility, launch CMake and build all the available examples.

\subsection{Experiment workflow}

All what is needed to run the full set of experiments is: \texttt{bash reproduce.sh}.
This script will take care of running in a replicated manner (5 times) all available examples (3) in all the available configurations (3), for a total of 5\*3\*3=45 runs.
To run the experiments across multiple servers the json config files must be modified (see \S\ref{ssec:config}). 

\subsection{Evaluation and expected results}

The mean execution time for each combination will be reported on the output, and logs will be saved for each experiment in the respective folder.
If the computational platforms are the same as the one reported in the paper, each process is allocated to a different computational node, and each process is assigned a different set of cores (in case of multiple processes on a single node), then the obtained results should be congruent with the one reported in the paper.

\subsection{Experiment customization}
\label{ssec:config}

To configure the experiments, the \texttt{.json} files describing the distributed configuration of the computations have to be personalised.
All the instructions for doing that are available on the \texttt{README.md} file.
Additionally, inside the reproduce script, the \texttt{MAX\_ITER} variable can be set to change the replica factor of the experiments.

\subsection{Notes}

The reviewers can find additional information on the FFL software in the official README.md file.

\end{document}